# Two-photon imaging through a multimode fiber

Edgar E. Morales-Delgado,[1]* Demetri Psaltis,[2] and Christophe Moser[1]

[1]Laboratory of Applied Photonics Devices, School of Engineering, École Polytechnique Fédéral de Lausanne (EPFL), Station 17, 1015, Lausanne, Switzerland
[2]Laboratory of Optics, School of Engineering, École Polytechnique Fédéral de Lausanne (EPFL), Station 17, 1015, Lausanne, Switzerland
*edgar.moralesdelgado@epfl.ch

**Abstract:** In this work we demonstrate 3D imaging using two-photon excitation through a 20 cm long multimode optical fiber (MMF) of 350 µm diameter. The imaging principle is similar to single photon fluorescence through a MMF, except that a focused femtosecond pulse is delivered and scanned over the sample. In our approach, focusing and scanning through the fiber is accomplished by digital phase conjugation using mode selection by time gating with an ultra-fast reference pulse. The excited two-photon emission is collected through the same fiber. We demonstrate depth sectioning by scanning the focused pulse in a 3D volume over a sample consisting of fluorescent beads suspended in a polymer. The achieved resolution is 1 µm laterally and 15 µm axially. Scanning is performed over an 80x80 µm field of view. To our knowledge, this is the first demonstration of high-resolution three-dimensional imaging using two-photon fluorescence through a multimode fiber.

**1. Introduction**

An endoscope allows the visualization of the interior of organs or cavities inside of the body. A widely used endoscopic system for clinical use is based on multi-core fibers (MCFs). The main benefit of a MCF is that it is 1. thin (0.3 mm for 6,000 cores) and 2. the part entering the body is passive, i.e does not contain active electronic components. However, MCF deliver pixelated images due to the gap between the cores (3-5 µm), which limits the resolution of the system to several micrometers [1-8].

In a similar fashion to scattering media [9-13], wavefront shaping methods using MCFs and multimode fibers (MMFs) have been able to effectively "de-pixelate" images to obtain diffraction limited resolution. Wavefront shaping methods include phase conjugation in the analog or digital domain [14-19], transmission matrix to provide wide field imaging [20-23] and iterative optimization [24-30]. However, these imaging approaches have been demonstrated using monochromatic light sources. Thus, for fluorescence imaging, only one-photon excitation was possible.

Two-photon excited fluorescence (TPEF) is obtained with pulsed light sources in the femtosecond regime. Andersen et al have demonstrated two-photon imaging through a custom MCF with 169 cores [31]. We have previously shown scanning and focusing of pulses in a step index MMF while maintaining a pulse duration of 500 fs through 30 cm fiber propagation [32]. In graded index fibers, spatial focusing of pulses has been shown [33]. However, in these previous studies TPEF imaging through multimode fibers was not demonstrated.

In this paper, we demonstrate, for the first time, high resolution two-photon excitation imaging through a graded-index multimode optical fiber. We use time-gated digital phase conjugation to focus and scan high intensity femtosecond pulses through the fiber [32]. Sectioned images of fluorescent beads suspended in a PDMS polymer volume are imaged. The high resolution (1µm), small probe diameter (350 µm), large collection efficiency, sectioning capability and ultrashort focused pulse delivery (120 fs) of our system opens the possibility for in-vivo minimally invasive multiphoton endoscopy in areas such as the brain, the ear or the eye.

## 2. Focusing pulses through the multimode fiber

*2.1 Calibration*

To generate spatially focused pulses through the multimode fiber, we use time-gated digital phase conjugation, which requires the following calibration step depicted in Fig.1. The 100 fs pulses from a Ti:sapphire laser are stretched out in time by a prism pair separated by 164 cm as described in Appendix A. This compensates for the group velocity dispersion (GVD) that the pulses suffer as they travel through the fiber. The pre-chirped beam (calibration beam) is then focused in front of the 20 cm length graded index fiber by a 40x microscope objective.. We call this the "calibration spot". As light from the calibration spot enters the MMF it is decomposed into multiple fiber modes, which generate, after propagation to the proximal side of the fiber, a temporally dispersed and spatially scrambled light field. In an off-axis holographic arrangement, a digital hologram of the time-gated light field is recorded on the CCD at the other end of the fiber. The process is repeated at different locations of the calibration spot in a three-dimensional grid. The digital holograms are stored. Recently, a rotational memory effect of multimode fibers has been reported, which can reduce the calibration time [34]. Fiber bending introduces changes in the propagation of the fiber modes. In our set-up, recalibration is required if the fiber is bent. However, this can be solved by incorporating a bending-compensation method via geometric characterization of fiber bending [20] or by measurement of the proximal speckle of a fiber with a distal reflective coating [35].

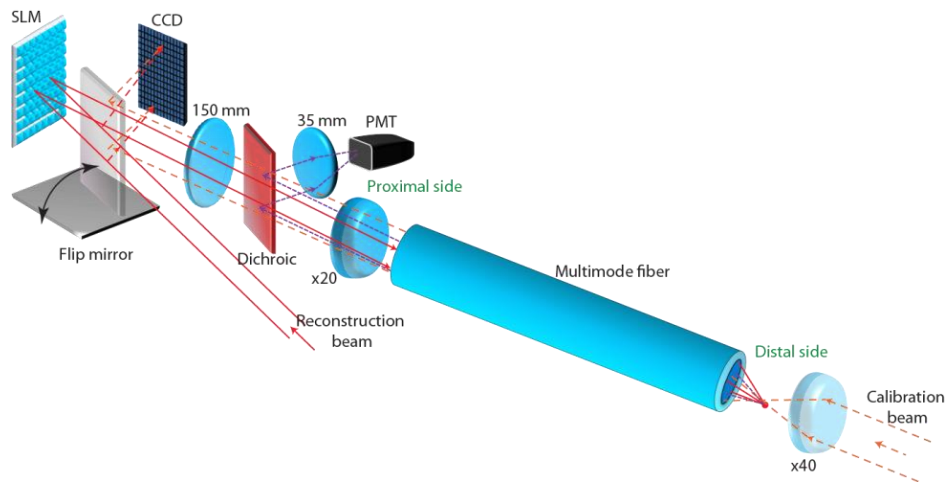

Fig. 1. (a) Experimental Setup. *Calibration step.* A calibration beam, previously pre-chirped, is focused and coupled into the multimode fiber. The field emerging from the proximal side of the fiber is interfered with the reference beam. An off-axis digital hologram is recorded. *Reconstruction step.* The flip mirror is removed. The reference reconstructs the phase conjugated version of the recorded hologram. Light counter propagates through the fiber, generating the initial focus on the distal side. *Imaging step.* The focused pulses are sequentially scanned over the specimen. The emitted two-photon fluorescence (2PF) is collected through the fiber and measured with the PMT.

*2.2 Digital scanning of pulsed light*

Once the calibration process is completed, the phase conjugated field of each recorded hologram is sequentially reconstructed (the flip mirror is removed) with a spatial light modulator (SLM). The reconstructed field counter propagates through the fiber generating a temporally and spatially sharp focused spot at the distal end. We define the enhancement of the focused pulse as the maximum intensity value in the focused spot divided by the average intensity in the background. As in scattering media, a larger number of controlled modes yields a better focusing capability; hence, greater enhancement [9]. Modes in graded index fibers are confined closer to the core and suffer less modal dispersion than in step-index fibers. However, this results in a drop of the enhancement value when the focus is away from the center of the fiber. Figure 2 shows the experimental results of intensity, enhancement, and spot size of a 40x40 grid of focused pulses generated sequentially 100 µm away from the facet of the fiber. The spots are spaced 2µm apart. For illustration purposes, Figure 2 (a) shows half of the generated spots (every 4 µm).

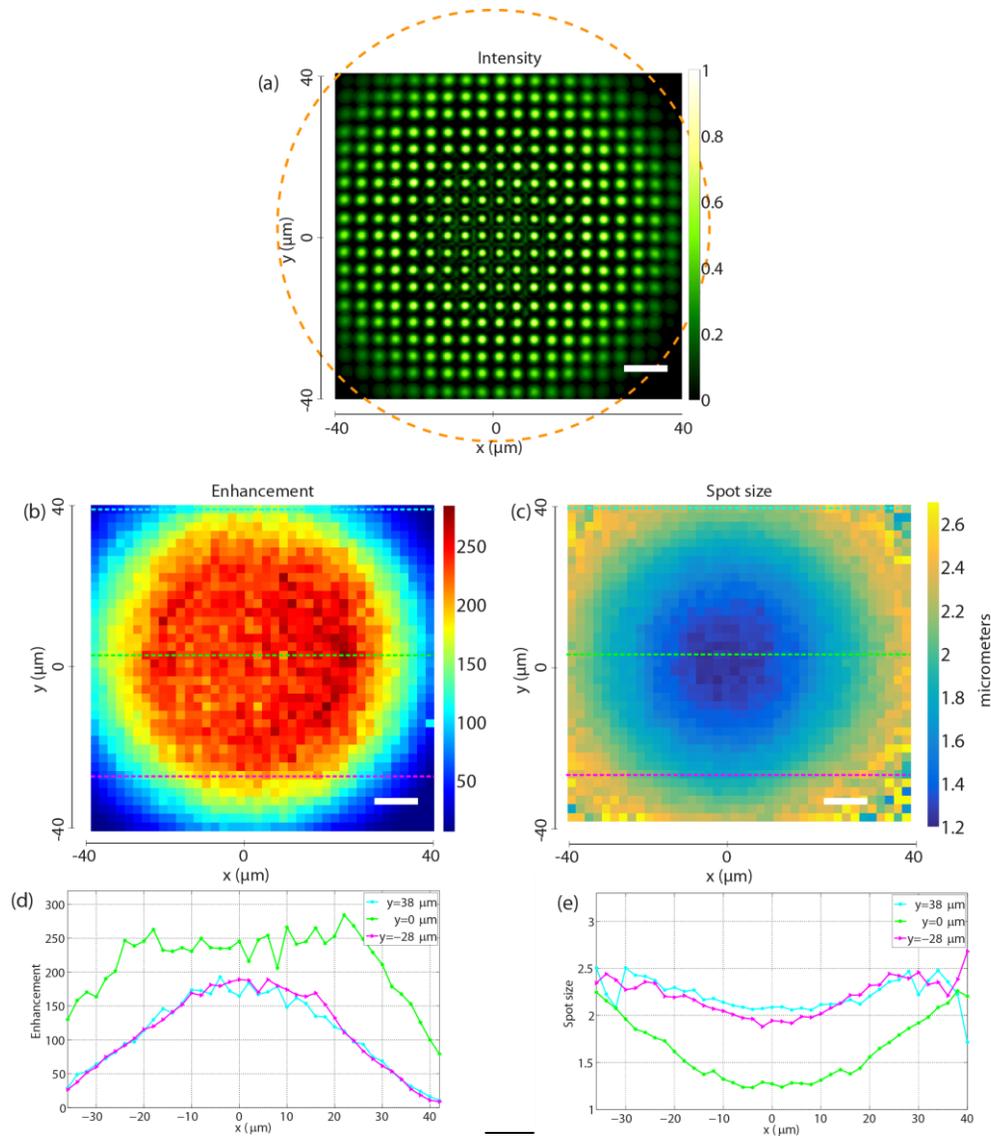

Fig. 2. Scanning of the focused spot at the fundamental wavelength of 800 nm. (a) Sequential superposition of a focused pulse scanned in a 80x80 µm field of view. Scanning step is 4 µm. (b) Intensity enhancement of the focused pulse as a function of scanning position. (c) FWHM size of the focused pulses as a function of scanning position. (d) Line profile of (b) at the location of the dashed lines. (e) Line profile of (c) at the location of the dashed lines. Scale bars are 10 µm. The dashed orange circle represents the core of the multimode fiber.

The FWHM spot size of the focused pulses is inversely proportional to the numerical aperture of the fiber. For an observer located on the distal side, the effective numerical aperture of the fiber depends on the position of observation, being maximum at the center [15]. Therefore, the spot size is minimum at focusing locations near the optical axis as shown in Figure 2 (c) and 2 (e).

*2.3 GVD compensation*

As demonstrated in [32], the counter-propagation of a time-sampled set of modes excites only fiber modes that travel at approximately the same speed (group velocity). These modes generate a focused pulse which does not suffer from modal dispersion. However, group velocity dispersion GVD still occurs. GVD is a phenomenon by which the spectral components of light travel through a material at different speeds ($dn/d\lambda \neq 0$) which results in broadening of the propagating pulse in time[36].

In our experiments, we compensate for the GVD occurring in the 20 cm length of the fiber by passing the beam through a prism pair (shown in Fig. 8) separated by 164 cm, which introduces a chirp and a total group delay of $-7200\ fs^2$. Dispersion compensation enhances the visibility of the fringes at the calibration step, and allows the reconstruction of a 120 fs pulse when used at the reconstruction step, close to the initial 100 fs. In our experimental setup, the compensation of GVD is optional at the reconstruction step. Figure 3 compares the intensity and the pulse length of a line of pulses focused through the fiber when the hologram is reconstructed with the un-chirped (no GVD compensation) and with the chirped (GVD compensation) reference. In the case of no GVD compensation, the pulse length is longer, on average 222 fs, due to the temporal broadening produced by GVD in the fiber (Fig. 3 (d)). In contrast, when the pulse is pre-chirped before reconstruction, the spectral dispersion of the excited modes is canceled. All spectral components arrive simultaneously at the distal side of the fiber and produce a short focused pulse, on average 117 fs. A small temporal broadening is observed, from 100 fs to 117 fs which is attributed to the GVD set-up where some spectral components angularly dispersed by the first prism (Fig. 8) are not collected by the second prism. The slight variations in pulse width (a standard deviation of 15 fs for no GVD compensation and 17 fs in the other case) are produced by slightly different propagation paths in the fiber, depending on the location of the generated spot and the specific excited modes. Figures 3 (e) and (f) show the second order interferometric autocorrelation trace and envelope of pulses focused at the center of the fiber. For the case of no GVD compensation (Fig 3(f)), we can notice an increased non-oscillating intensity at the tails of the pulse, which indicates that the pulse is chirped. For the case of GVD compensation (Fig 3(e)), the effect is less evident because material dispersion in the fiber is compensated. More details on the pulse width measurements can be found in Appendix A. Regarding the intensity enhancement, it depends on the position of the focused pulse as shown in Fig. 2 (b) and (d). The enhancement of the generated focused spots (Fig 3 (c)) is on average 252 and 220 within a 40 $\mu m$ radius from the optical axis, for the case of GVD and no GVD compensation, respectively. Hence, the spectral correction achieved by pre-chirping the reconstructed pulse not only reduces pulse broadening, but also enhances the intensity of the focused spot. Since the cross section of the optical fiber is azimuthally symmetric, for the described results, we can characterize the spots scanned only along a line passing through the optical axis at the center of the core without loss of generality. Taking into account a transmittance of 75% through the GVD compensation set-up (prism pair), the signal to noise ratio of the focused spot is enhanced by a factor of 1.6 when GVD is compensated.

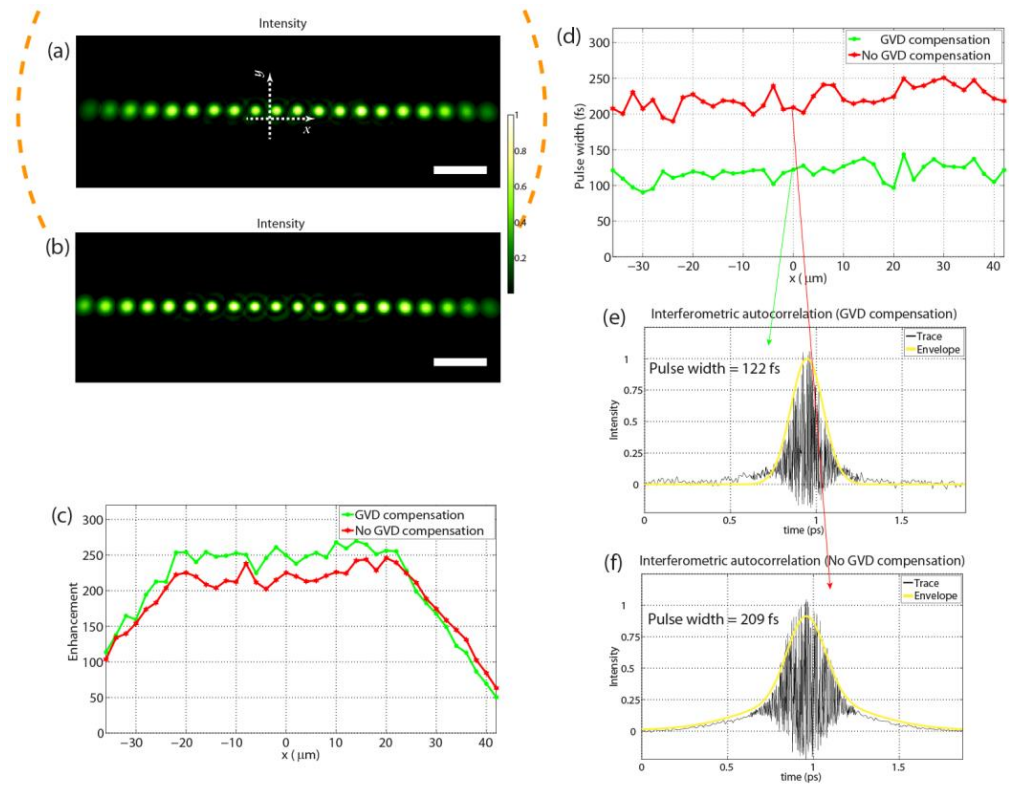

Fig. 3. Pulse width characterization. Intensity of the generated focused pulses at 800 nm without (a) and with (b) GVD compensation. (c) Intensity enhancement vs position. (d) Pulse width vs position. (e), (f) Sample interferometric autocorrelation traces of the central focused pulses with and without GVD compensation respectively. Scale bars are 10 µm. The dashed orange semi-circle represents the core of the multimode fiber.

### 3. Two-photon imaging through the multimode fiber

Once we are able to generate and scan a grid of focused pulses at different planes through the multimode fiber, the next step is to use the system as a two-photon endoscope. In 2PFE, two photons of long excitation wavelength are simultaneously absorbed by a fluorophore. When the fluorophore returns from the excited to the ground state, a photon of higher energy than any of the two absorbed photons is emitted [37]. We prepared a sample of 1-5 µm fluorescent red microspheres suspended in a transparent PDMS volume of 55 µm thickness. We used an excitation wavelength of 800 nm. The spectral emission of the beads is centered at 575 nm. For more details about the sample preparation, see Appendix A. We scan the focused pulse every 700 nm in a grid of 40x70 points (28x49 µm FOV). We scan the planes from 0 to 50 µm inside the sample. Figure 4 shows the sectioned images.

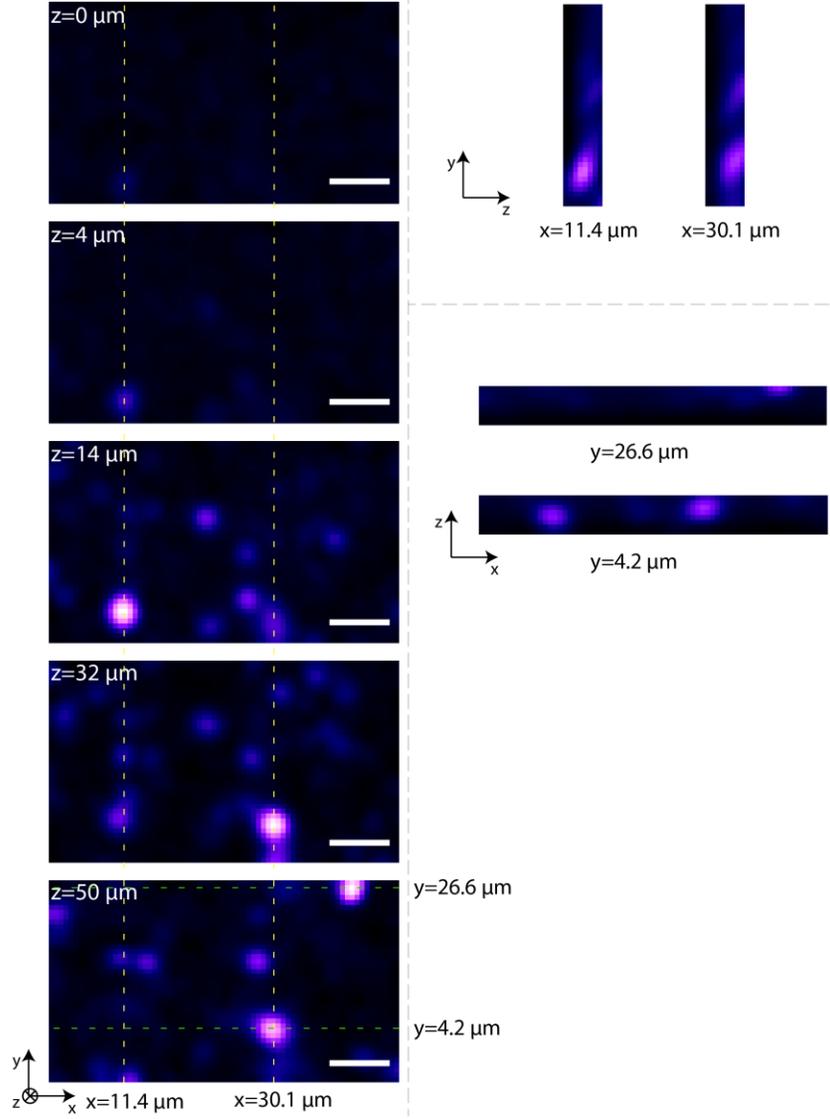

Fig. 4. Two-photon images of fluorescent beads. (y,x,z) Field of view is 28x49x50 µm. Scale bars are 10 µm. To view all the frames see Media 1.

The number of fluorescent photons produced in two-photon emission is given by [37]:

$$n_a \approx \frac{p_0^2 \delta}{\tau_p f_p^2} \left(\frac{NA^2}{2hc\lambda}\right)^2$$

Where $n_a$ is the number of fluorescent photons, $p_0$ is the average power, $\tau_p$ is the pulse length, $f_p$ is the repetition rate of the source, $\delta$ is the two-photon absorption cross section and NA is the numerical aperture of the fiber. As described before, the intensity enhancement and spot size is a function of the scanning position at the distal end of the fiber. Hence, we calculate a quantity proportional to $n_a$ called the normalized two-photon fluorescence (2PF normalized), which is shown in Fig. 5. In a homogeneous fluorescent medium, focused pulses generated near the center of the optical axis produce a larger number of fluorescent photons. Therefore, to reconstruct correctly the original image, we require a compensation of the over-estimation of intensity in the central area, which is performed as a post-acquisition process.

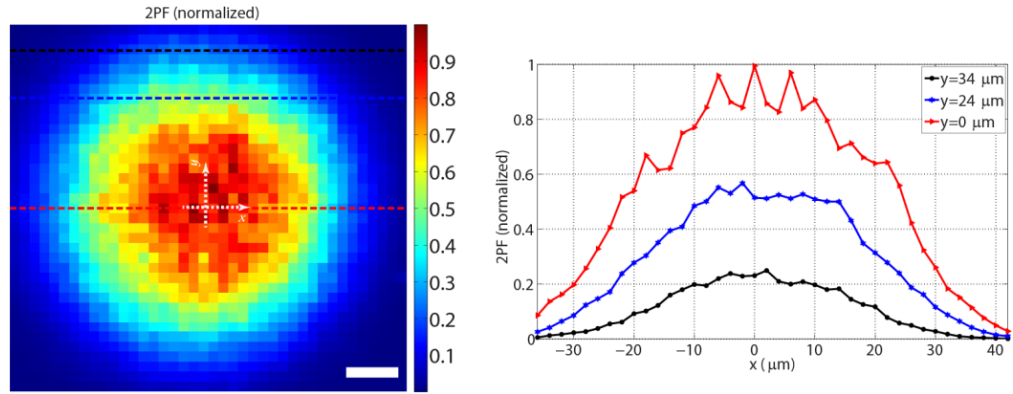

Fig 5. Normalized fluorescent photons as a function of position. Scale bar is 10 µm

Two-photon fluorescence occurs in a smaller volume than single photon absorption. Hence, the point spread function (PSF) of the two-photon imaging fiber device is $\sqrt{2}$ times smaller than the linear PSF. Additionally, 2PF suppresses the out of focus excitation. This is illustrated in Fig. 6, which shows the linear and 2PF PSF of the fiber imaging system. The intensity after the first zero of the airy disk is significantly reduced by the two-photon process as seen in Fig. 6. In summary, the 2PF images acquired with the proposed system has a resolution smaller than 1.05 µm within a 40 µm radius from the optical axis with a two-photon equivalent enhancement larger than 57,000. The maximum average power delivered to the sample is 22 mW, which corresponds to a spot energy of 0.27 $mJ/cm^2$.

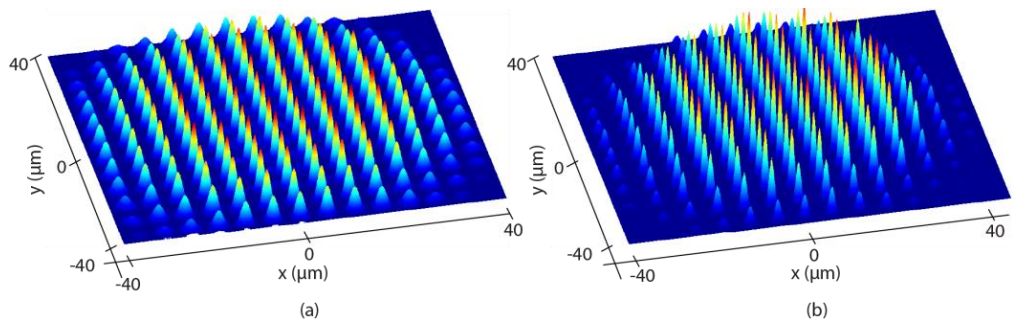

Fig. 6. Experimental PSF of the multimode fiber imaging system. (a) Single-photon fluorescence. (b) Two-photon fluorescence.

## 4. Conclusions

We have demonstrated the first implementation of a multimode fiber two-photon endoscope. Our digital scanning method does not require distal mechanical elements, allowing an ultra-thin probe of 350 µm diameter. The large area of the fiber and high numerical aperture (0.34 NA) allows large collection efficiency. The reduced modal dispersion of graded-index fibers allows the simultaneous control of a larger number of modes, generating a brighter focused pulse compared to the case of step-index fibers. Time-gated phase conjugation of a pre-chirped beam prevents pulse broadening due to GVD of the pulse propagating through the fiber, generating an ultrashort pulse of 120 fs duration. The high intensity focused pulse is generated outside of the fiber avoiding nonlinearities in the propagating pulse. Our work sets the basis for in-vivo ultra-thin two-photon endoscopy.

## Appendix A: Materials and methods

The experiments were conducted using a Ti:sapphire oscillator (Coherent Mira 900; central wavelength $\lambda_c$=800 nm; spectral width $\sigma_\lambda$=10 nm; pulse width=100 fs). The capability of our method to form a sharp focused pulse through the fiber depends on the number of modes than can be controlled simultaneously [32]. In step-index fibers, the pulse spreading in time is given by: $\sigma_\tau \approx L\Delta/(2c_1)$ while in graded-index fibers: $\sigma_\tau \approx L\Delta^2/(4c_1)$ [36]. Due to their smaller modal temporal-spreading, we selected a graded-index fiber for our experiments. The fiber has a 200 µm core diameter, 280 µm cladding diameter, 20 cm length, 0.29 NA and supports 26,000 modes at a 800 nm wavelength. The SLM used for time-gated phase conjugation is a Holoeye Pluto, 1920x1080 pixels, phase only.

To increase the resolution of the two-photon imaging system a grin lens (0.5 mm length, 70° view angle, 250 µm diameter) is attached to the distal end of the multimode fiber. (Fig. 7) The lens increases the numerical aperture of the fiber from 0.29 to 0.34 (resolution changed from 1.2 µm to 1 µm) but reduces the circular field of view of the fiber from 200 µm diameter to 106 µm. The multiphoton imaging principle demonstrated in this work can be extended to lensless and thinner graded index fibers without loss of generality.

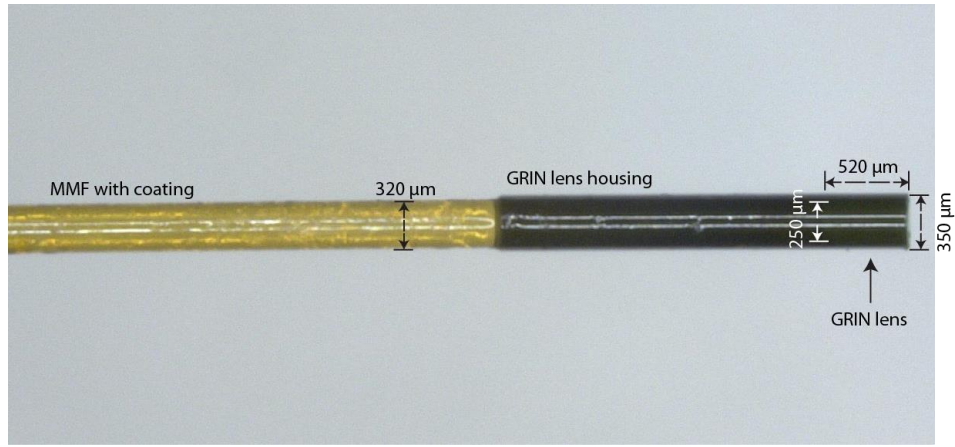

Fig. 7. Multimode fiber probe. The white horizontal lines are the reflection of the lamp of the microscope.

At the calibration step, (Fig. 8, flip mirror 1 and 2 are removed) the calibration beam passes through the prism pair and is coupled into the fiber by a 40x microscope objective (OBJ2). The field emerging from the fiber on the proximal side is imaged by a 20x objective (OBJ1) and a 150 mm lens (L1) onto the CCD, where it is interfered by the pulsed reference beam. The pulsed reference can be delayed and is set at the time in which the visibility of the fringes is maximum [32].

At the reconstruction step, flip mirror 3 is removed allowing the reference to reach the SLM and reconstruct the phase conjugate version of the recorded field. Light counter propagated through the fiber and is focused on the distal side, and imaged on Camera 2 by a 200 mm lens (L2). The hologram can be reconstructed with GVD compensation (flip mirrors 1 and 2 in place) and without GVD compensation (flip mirrors 1 and 2 removed).

At the imaging step, the 2PF is collected through the fiber, reflected by a 650 nm short-reflected dichroic mirror, and focused on the PMT by a 35 mm lens (L3). This ensures the collection of most of the spectral emission of the sample, whose maximum fluorescence is centered at 575 nm.

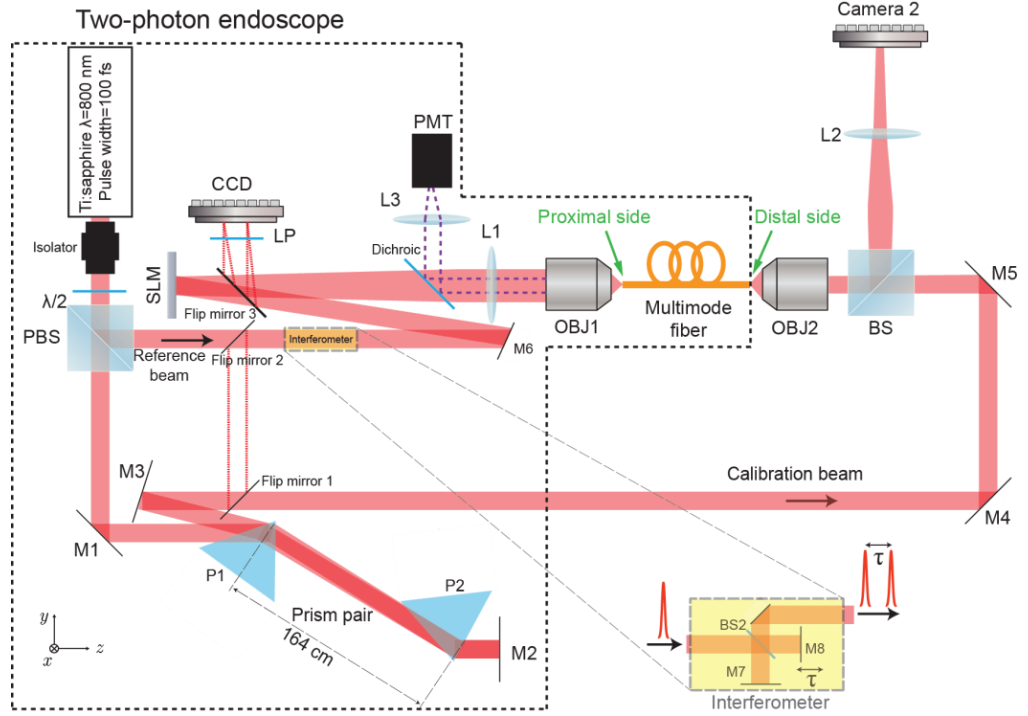

Fig. 8 Complete diagram of the experimental setup.

To measure the pulse width of the generated pulses through the fiber we use second order interferometric autocorrelation by introducing two replicas of the pulses on the reference arm using the interferometer shown in Figure 8. As a nonlinear element for the second order interferometer, we place a homogeneous sample of PDMS with Rhodamine 6G [38, 39]. The fluorescence is collected through the fiber and measured by the PMT. The background to maximum ratio of the measured second order interferometric traces is 8. A $sech^2$ pulse is assumed to calculate the pulse lengths (0.64 times the width of the envelope of the autocorrelation trace).

The image sample used in these experiments consists of 1-5 μm fluorescent beads (Cospheric FMR – Red Fluorescent Microspheres) suspended in a transparent PDMS volume. The spectral emission of the beads is centered at 575 nm. The fluorescent beads were mixed with PDMS in a concentration of 1 mg of beads per 100 mg of PDMS with 10 mg of curing agent. A drop of the mixture was placed on a 100 μm thick glass slide and spin coated 60 seconds at 1000 rpm to achieve a flat 55 μm PDMS thickness. Finally, the sample was cured in a hot plate at 80° C for 60 minutes.

To compensate the GVD of the fiber, we use a prism pair in a reflection configuration (Fig. 8). For their high angular dispersion and low loss, we selected prisms made of SF10 of 50x50x50 mm. The separation between the prisms was calculated using the following group delay dispersion (GDD) equation for a prism pair [40]:

$$GDD_{prism\,pair} = \frac{\lambda^3}{2\pi c^2}\left[4l\left\{\left[\frac{d^2n}{d\lambda^2} + \left(2n - \frac{1}{n^3}\right)\left(\frac{dn}{d\lambda}\right)^2\right]sin\beta - 2\left(\frac{dn}{d\lambda}\right)^2 cos\beta\right\} + 8D\left(\frac{d^2n}{d\lambda^2}\right)\right]$$

Where $n$ is the refractive index of SF10, D is the beam diameter and $\beta = -2\frac{dn}{d\lambda}\Delta\lambda$.

The prisms themselves introduce a GDD of the same sign as the GVD introduced by the fiber. To compensate that, the separation between the prisms is larger than that of a prism pair made of infinitesimally small prisms. For adjustment, the second prism was mounted in a 2D stage that modes parallel and perpendicular to the beam connecting prism 1 (P1) and prism 2 (P2).

## 5. Acknowledgments

We would like to thank Dr. Carlos Macias-Romero from the Laboratory for Fundamental BioPhotonics at EPFL, for his helpful discussions about microscopy and nonlinear optics.